%% file: adaptmul.tex
\documentclass{article}
\pdfoutput=1

\usepackage{natbib}

\usepackage{amsmath}
\usepackage{amssymb}
\usepackage{amsthm}
\usepackage[svgnames]{xcolor}
\usepackage{url}
\usepackage[ruled,vlined,linesnumbered,algo2e]{algorithm2e}

\usepackage[pdfauthor={Daniel S. Roche},%
            pdftitle={Chunky and Equal-Spaced Polynomial Multiplication},%
	         pdfsubject={Symbolic Computation},%
            colorlinks,linkcolor=DarkBlue,citecolor=DarkGreen,%
           pdfkeywords={Polynomial multiplication, adaptive algorithms, %
           sparse polynomials}]{hyperref}

\DeclareMathOperator{\loglog}{loglog}
\DeclareMathOperator{\lcm}{lcm}

\newcommand{\M}{\mathsf{M}}
\newcommand{\R}{\mathsf{R}}
\newcommand{\NN}{\mathbb{N}}
\newcommand{\ZZ}{\mathbb{Z}}

\theoremstyle{plain}
  \newtheorem{thm}{Theorem}[section]
  \newtheorem{lem}[thm]{Lemma}
  \newtheorem{cor}[thm]{Corollary}

\newenvironment{pf}{\begin{proof}}{\end{proof}}
\newenvironment{ack}{\section*{Acknowledgements}}{}

\numberwithin{equation}{section}

\definecolor{DarkGreen}{rgb}{0,.35,0.2}
\definecolor{DarkBlue}{rgb}{0,0,.545}

\begin{document}

\title{Chunky and Equal-Spaced\\Polynomial Multiplication}

\author{Daniel S. Roche\\[4mm]
\begin{minipage}{\textwidth}
\centering\footnotesize
Symbolic Computation Group \\
Cheriton School of Computer Science \\
University of Waterloo \\
Waterloo, Ontario, Canada \\
\texttt{droche@cs.uwaterloo.ca} \\
\url{http://www.cs.uwaterloo.ca/~droche/}
\end{minipage}}

\maketitle

\begin{abstract}
Finding the product of two polynomials is an essential and basic problem in
computer algebra. While most previous results have focused on the worst-case
complexity, we instead employ the technique of adaptive analysis to give an
improvement in many ``easy'' cases.
We present two adaptive measures and methods for polynomial
multiplication, and also show how to effectively combine them to gain
both advantages. One useful feature of these algorithms is that they
essentially provide a gradient between existing ``sparse'' and ``dense''
methods. We prove that these approaches provide significant improvements
in many cases but in the worst case are still comparable to the fastest
existing algorithms.
\end{abstract}

\input{adaptmul-body}

\bibliographystyle{plainnat}
\bibliography{adaptmul}

\end{document}

%% file: adaptmul-body.tex
\section{Introduction}
Computing the product of two polynomials is one of the most important
problems in symbolic computation, and the operation is part of the basic
functionality of any computer algebra system. 
We introduce new multiplication algorithms which use the technique of
adaptive analysis to gain improvements compared to existing approaches both in
theory and in practice.

\subsection{Background}

For what follows, $\R$ is an arbitrary ring (commutative, with
identity), such that ring elements have unit storage and basic ring
operations have unit cost. In complexity estimates, we also count operations
on \emph{word-sized integers}, which are assumed only to be large enough (in
absolute value) to store the size of the input. 

There are essentially two representations for univariate polynomials over $\R$,
and existing algorithms for multiplication require one of these representations.
Let $f\in\R[x]$ with degree less than $n$ written as
\begin{equation}
f = c_0 + c_1 x + c_2 x^2 + \cdots + c_{n-1} x^{n-1}, \label{eqn:dense} 
\end{equation}
for $c_0,\ldots,c_{n-1}\in\R$. 
The \emph{dense representation} of $f$ is simply
an array $[c_0,c_1,\ldots,c_{n-1}]$ of
length $n$. 

Next, suppose that at most $t$ of the coefficients are nonzero, so
that we can write
\begin{equation}
f = a_1 x^{e_1} + a_2 x^{e_2} + \cdots + a_t x^{e_t},
\label{eqn:sparse}
\end{equation}
for $a_1,\ldots,a_t\in\R$ and $0\leq e_1 < \cdots < e_t$.
Hence $a_i = c_{e_i}$ for $1\leq i \leq t$, and in particular $e_t = \deg f$.
The \emph{sparse representation} of $f$ is a list of coefficient-exponent
tuples $(a_1,e_1),\ldots,(a_t,e_t)$. The exponents in this case could be
multi-precision integers, and so the total size of the sparse representation is
proportional to $\sum_i (1 + \log_2 e_i)$. This is bounded below by
$\Omega(t\log t + \log n)$ and above by $O(t\log n)$.

Algorithmic advances in dense polynomial multiplication have generally followed
results for long integer multiplication. The $O(n^2)$ school method 
was first improved by \citet{KarOfm63} to $O(n^{1.59})$ with
a two-way divide-and-conquer scheme, later generalized to $k$-way by
\citet{Too63} and \citet{Coo66}. 
\citet{SchStr71} developed the first pseudo-linear time algorithm for
integer multiplication with cost $O(n\log n\loglog n)$;
this is also the cost of the fastest known algorithm for polynomial
multiplication \citep{CanKal91}. 

In practice, all of these algorithms will
be used in certain ranges, and so we employ the usual notation of a
\emph{multiplication time} function $\M(n)$, 
the cost of multiplying two dense polynomials with degrees both less than
$n$. Also
define 
$\delta(n)=\M(n)/n$.
If $f,g\in\R[x]$ with different degrees
$\deg f<n$, $\deg g<m$, and $n>m$, by 
splitting $f$ into $\lceil n/m \rceil$ size-$m$ blocks we can
compute the product $f\cdot g$ with cost $O(\frac{n}{m}\M(m))$, or
$O(n\cdot\delta(m))$.

For the multiplication of two sparse polynomials as in \eqref{eqn:sparse},
the school method uses $O(t^2)$ ring operations, which cannot be improved
in the worst case. However, since the degrees could be very large,
the cost of exponent arithmetic becomes significant.
The school method uses $O(t^3\log n)$ word operations and $O(t^2)$ space.
\citet{Yan98} reduces the number of word operations to $O(t^2 \log t \log n)$
with the ``geobuckets'' data structure. Finally, recent work by
\citet{MonPea07}, following \citet{Joh74}, gets this same time complexity but
reduces the space requirement to $O(t+r)$, where $r$ is the number of nonzero
terms in the product.

The algorithms we present are for univariate polynomials. They can also be used
for multivariate polynomial multiplication by using 
\emph{Kronecker substitution}:
Given two $n$-variate polynomials $f,g\in\R[x_1,\ldots,x_n]$ with max degrees
less than $d$, substitute $x_i = y^{(2d)^{i-1}}$ for 
$1\leq i\leq n$, multiply the univariate polynomials over $\R[y]$, then convert
back. Many other representations exist for multivariate polynomials
\citep[see][]{Fat02}, but we will not compare with them or consider them
further.

\subsection{Overview of Approach}

The performance of an 
\emph{adaptive algorithm} depends not only on the
size of the input but also on some inherent difficulty measure.
Such algorithms match standard approaches in their worst-case performance,
but perform far better on many instances. This idea was first applied to
sorting algorithms and has proved useful both in theory and in practice
\citep[see][]{PetMof95}. Such techniques have also proven
useful in symbolic computation, for example the early termination
strategy of \citet{KalLee02}.

\emph{Hybrid algorithms} combine multiple different approaches to the same
problem to effectively handle more cases \citep[e.g.][]{DurSauWan03}.
Our algorithms are also hybrid in the sense that they provide a smooth
gradient between existing sparse and dense multiplication algorithms.
The adaptive nature of the algorithms means that in fact they will be faster
than existing algorithms in many cases, while never being (asymptotically)
slower.

The algorithms we present will always proceed in three stages. First, the
polynomials are read in and converted to a different representation which
effectively captures the relevant measure of difficulty. Second, we multiply the
two polynomials in the alternate representation. Finally, the product is
converted back to the original representation. 

The computational cost of the second step (where the multiplication is
actually performed) depends on the difficulty of the particular instance.
Therefore this step should be the dominating cost of the entire algorithm,
and in particular
the cost of the conversion steps must be linear in the
size of the input polynomials.

In Section~\ref{sec:chunky}, we give the first idea for adaptive
multiplication, which is to write a polynomial as a list of dense
``chunks''. The second idea, presented in Section~\ref{sec:eqspace},
is to write a polynomial with ``equal spacing'' between coefficients as
a dense polynomial composed with a power of the indeterminate.
Section~\ref{sec:combine} shows how to combine these two ideas to make
one algorithm which effectively captures both difficulty measures.
Finally, a few conclusions and ideas for future directions are discussed
in Section~\ref{sec:conclusion}.

Preliminary progress on some of these results was presented at the Milestones
in Computer Algebra (MICA) conference held in Tobago in May
2008 \citep{Roc08}.

\section{Chunky Polynomials}
\label{sec:chunky}

The basic idea of chunky multiplication
is a straightforward combination of the standard sparse and dense
representations, providing a natural gradient between the two approaches for
multiplication. We note that a similar idea was noticed (independently)
around the same time by \citet[page 11]{Fat08}, although the treatment here
is much more extensive.

For $f \in \R[x]$ of degree $n$, the \emph{chunky representation}
of $f$ is a sparse polynomial with
dense polynomial ``chunks'' as coefficients:
\begin{equation} \label{eqn:chunkyrep}
f = f_1 x^{e_1} + f_2 x^{e_2} + \cdots + f_t x^{e_t},
\end{equation}
with $f_i \in \R[x]$ and $e_i \in \NN$ for each $1\leq i\leq t$.
We require only that
$e_{i+1} > e_i+\deg f_i$ for $1\leq i\leq t-1$, and each $f_i$ has
nonzero constant coefficient.

Recall the notation introduced above of $\delta(n)=\M(n)/n$.
A unique feature of our approach is that we will
actually use this function to tune the algorithm. 
That is, we assume a subroutine is given to evaluate $\delta(n)$ for any
chosen value $n$.

If $n$ is a word-sized integer, then the computation of $\delta(n)$ must
use a constant number of word operations. If $n$ is more than
word-sized, then we are asking about the cost of multiplying two dense
polynomials that cannot fit in memory, so the subroutine should return
$\infty$ in such cases.
Practically speaking, the $\delta(n)$ evaluation will
usually be an approximation of the actual value, but for what follows we
assume the computed value is always exactly correct. 

Furthermore, we
require $\delta(n)$ to be an increasing function which grows more
slowly than linearly, meaning that for any $a,b,d\in \NN$ with $a<b$,
\begin{equation}\label{eqn:deltacond}
\delta(a+d) - \delta(a) \geq \delta(b+d)-\delta(b).
\end{equation}
These conditions are clearly satisfied for
all the dense multiplication algorithms and corresponding $\M(n)$ functions
discussed above, including the piecewise function used in practice.

The conversion of a sparse or dense polynomial to the chunky
representation proceeds in two stages: first, we compute an ``optimal
chunk size'' $k$, and then we use this computed value as a parameter in
the actual conversion algorithm. The product of the two polynomials is
then computed in the chunky representation, and finally the result is
converted back to the original representation.
The steps are presented in reverse order in the hope that the goals at
each stage are more clear.

\subsection{Multiplication in the chunky representation}

Multiplying polynomials in the chunky representation uses sparse multiplication
on the outer loop, treating each dense polynomial chunk as a coefficient, 
and dense
multiplication to find each product of two chunks. 

For $f,g\in\R[x]$ to be multiplied, write $f$ as in \eqref{eqn:chunkyrep}
and $g$ as
\begin{equation}\label{eqn:chunkyg}
g = g_1x^{d_1} + g_2x^{d_2} + \cdots + g_{s}x^{d_{s}},
\end{equation}
with $s\in\NN$ and similar conditions on each $g_i\in\R[x]$ and $d_i\in\NN$
as in \eqref{eqn:chunkyrep}. Without loss of generality, assume also that
$t\geq s$,
that is, $f$ has more chunks than $g$.
To multiply $f$ and $g$, we need to compute each product $f_ig_j$ 
for $1\leq i\leq t$ and $1\leq j\leq s$ and
put the resulting chunks into sorted order. It is likely that some of
the chunk products will overlap, and hence some coefficients
will also need to be summed.

By using heaps of pointers as in \cite{MonPea07}, the chunks of the
result are computed in order, eliminating unnecessary additions and
using little extra space. 
A min-heap of size $s$ is filled
with pairs $(i,j)$, for $i,j\in\NN$,
and ordered by the corresponding sum of
exponents $e_i+d_j$. 
Each time we compute a new chunk product $f_i\cdot g_j$,
we check the new exponent against the degree of the previous chunk,
in order to determine whether to make a new chunk in the product
or add to the previous one. The details of this approach are given
in Algorithm~\ref{alg:chunkymul}.

\begin{algorithm2e}[htb]
\DontPrintSemicolon
\KwIn{$f,g$ as in \eqref{eqn:chunkyrep} and \eqref{eqn:chunkyg}}
\KwOut{The product 
$f\cdot g=h$ in the chunky representation}
\medskip
$\alpha \gets f_1\cdot g_1$ using dense multiplication \;
$b \gets e_1 + d_1$\;
$H \gets$ min-heap with pairs 
$(1,j)$ for $j=2,3,\ldots,s$, 
ordered by exponent sums \;
\lIf{$i \geq 2$}{insert $(2,1)$ into $H$}\;
\While{$H$ is not empty}{
	$(i,j) \gets$ pair from top of $H$\label{alg:chunkymul:heapout}\;
  $\beta \gets f_i\cdot g_j$ using dense multiplication 
		\label{alg:chunkymul:densemul}\;
  \If{$b + \deg \alpha < e_i + d_j$}{
    \textbf{write} $\alpha x^b$ as next term of $h$\;
    $\alpha \gets \beta;\quad b \gets e_i+d_j$
  }
  \lElse{$\alpha \gets \alpha + \beta x^{e_i+d_j-b}$
  stored as a dense polynomial\label{alg:chunkymul:adds}}\;
	\lIf{$i<t$}{
	insert $(i+1,j)$ into $H$\label{alg:chunkymul:heapin}\;
  }
}
\textbf{write} $\alpha x^b$ as final term of $h$
\caption{Chunky Multiplication}
\label{alg:chunkymul}
\end{algorithm2e}

After using this algorithm to multiply $f$ and $g$, we can easily
convert the result back to the dense or sparse representation in linear
time. In fact, if the output is dense, we can preallocate space for the
result and store the computed product directly in the dense array, requiring
only some extra space for the heap $H$ and
a single intermediate product $h_{\rm new}$.

\begin{thm} \label{thm:chunkycost}
Algorithm~\ref{alg:chunkymul} correctly computes the product of $f$ and
$g$ using 
\[O\bigg(\sum_{\substack{\deg f_i \geq \deg g_j\\
1\leq i\leq t,\ 1\leq j\leq s}}(\deg f_i)
\cdot\delta(\deg g_j)
\quad+\quad\sum_{\substack{\deg f_i < \deg g_j\\
1\leq i\leq t,\ 1\leq j\leq s}}(\deg g_j)\cdot\delta(\deg f_i) \bigg)\]
ring operations and $O(ts\cdot\log s\cdot\log(\deg fg))$ word
operations.
\end{thm}
\begin{pf}
Correctness is clear from the definitions. The bound on ring operations
comes from Step~\ref{alg:chunkymul:densemul} using the fact that
$\delta(n) = \M(n)/n$. The cost of additions on
Step~\ref{alg:chunkymul:adds} is linear and hence also within the stated
bound.

The cost of word operations is incurred in removing from and
inserting to the heap on Steps \ref{alg:chunkymul:heapout} and
\ref{alg:chunkymul:heapin}. Because these steps are executed no more
than $t_ft_g$ times, the size of the heap is never more than $t_g$, 
and each exponent sum is
bounded by the degree of the product, the stated bound is correct.
\end{pf}

Notice that the cost of word operations is always less than the cost
would be if we had multiplied $f$ and $g$ in the standard sparse
representation. 
We therefore focus only on minimizing the number of ring
operations in the conversion steps that follow.

\subsection{Conversion given optimal chunk size}

The general chunky conversion problem is, given $f,g\in\R[x]$, both
either in the sparse or dense representation, to determine
chunky representations for $f$ and $g$ which minimize the cost of
Algorithm~\ref{alg:chunkymul}. Here we consider a simpler problem,
namely determining an optimal chunky representation for $f$ given that
$g$ has only one chunk of size $k$.

The following corollary comes directly from Theorem~\ref{thm:chunkycost}
and will guide our conversion algorithm on this step.
\begin{cor}
Given $f\in\R[x]$ as in \eqref{eqn:chunkyrep}, the number of ring
operations required to multiply
$f$ by a single dense polynomial with degree less than $k$ is
\[O\bigg( \delta(k) \sum_{\deg f_i\geq k}\deg f_i \quad + \quad
	k \sum_{\deg f_i < k} \delta(\deg f_i) \bigg)\]
\end{cor}

For any high-degree chunk (i.e. $\deg f_i\geq k$), we see that there
is no benefit to making the chunk any larger, as
the cost is proportional to the sum of the degrees of these
chunks. In order to minimize the cost of multiplication, then, we should
not have any chunks with degree greater than $k$ (except possibly in the
case that every coefficient of the chunk is nonzero), and we should
minimize $\sum \delta(\deg f_i)$ for all chunks with size less than $k$.

These observations form the basis of our approach in
Algorithm~\ref{alg:optconv} below. For an input polynomial $f\in\R[x]$,
each ``gap'' of
consecutive zero coefficients in $f$ is examined, in order. We determine
the optimal chunky conversion if the polynomial
were truncated at that gap. This is
accomplished by finding the previous gap of highest degree 
that should be
included in the optimal chunky representation. We already have the
conversion for the polynomial up to that gap (from a previous step), so
we simply add on the last chunk and we are done. At the end, after all
gaps have been examined, we have the optimal conversion for the entire
polynomial.

Let $a_i,b_i \in \ZZ$ for $0\leq i \leq m$ be the sizes of
each consecutive ``gap'' of zero coefficients and ``block'' of nonzero
coefficients, in order. Each $a_i$ and $b_i$ will be nonzero except possibly
for $a_0$ (if $f$ has a nonzero constant coefficient), and 
$\sum_{0\leq i \leq m} (a_i+b_i) = \deg f + 1$. For example, the
polynomial
\[f = 5x^{10} + 3x^{11} + 9x^{13} + 20x^{19} + 4x^{20} +
8x^{21}\]
has $a_0=10,b_0=2,a_1=1,b_1=1,a_2=5,$ and $b_2=3$. Also define $d_i$ to
be the degree of the polynomial up to (not including) gap $i$,
i.e. $d_i = \sum_{0\leq j < i} (a_j + b_j)$.

For the gap at index $\ell$, for $1\leq \ell\leq m$, we store the optimal chunky
conversion of $f \bmod x^{d_\ell}$ by a linked
list of indices of all gaps in $f$ that should also be gaps between
chunks in the optimal chunky representation. 
In $c_\ell$ we also store $1/k$ times the cost, in ring operations, of
multiplying $f \bmod x^{d_\ell}$ 
(in this optimal representation) by a single
chunk of size $k$. 

When examining the gap at index $\ell$, in order to determine the
previous gap of highest degree
to be included in the optimal chunky representation
if the polynomial were truncated at gap $j$, 
we need to find the index $i<\ell$ that minimizes
$c_i + \delta(d_\ell-d_i)$ (indices $i$ where
$d_\ell-d_i>k$ need not be considered, as discussed above). 
From
\eqref{eqn:deltacond}, we know that, if $1\leq i < j < \ell$ and
$c_{i}+\delta(d_\ell-d_{i})<c_{j}+\delta(d_\ell-d_{j})$,
then this same inequality continues to hold as $\ell$ increases.
That is, as soon
as an earlier gap results in a smaller cost than a later one, that
earlier gap will continue to beat the later one.

Thus we can essentially precompute the values of
$\min_{i<\ell} (c_i + \delta(d_\ell-d_i))$ by maintaining a stack of
index-index pairs. A pair $(i,j)$ of indices indicates that
$c_{i}+\delta(d_\ell-d_{i})$ is minimal as long as $\ell \leq j$. The
second pair of indices indicates the minimal value from gap $j$ to
the gap of the second index of the second pair, and so
forth up to the bottom of the stack and the last gap.

The details of this rather complicated algorithm are given in 
Algorithm~\ref{alg:optconv}.

\begin{algorithm2e}[htb]
\DontPrintSemicolon
\KwIn{$k\in \NN$, $f\in\R[x]$, and integers $a_i,b_i,d_i$ 
for $i = 0,1,2,\ldots,m$ as above}
\KwOut{A list $L$ of the indices of gaps to include in the optimal
chunky representation of $f$ when multiplying by a single chunk of size
$k$}
$L_1 \gets 0; \quad c_1 \gets \delta(b_0); \quad S \gets (0,m+1)$ \;
\For{$\ell=2,3,\ldots, m+1$}{
	\While{top pair $(i,j)$ from $S$ satisfies $j<\ell$ or
	$d_\ell-d_{i}>k$\label{alg:optconv:remold}}{
		Remove $(i,j)$ from $S$\;
	}
	\If{top pair $(i,j)$ from $S$ satisfies
		$c_{i}+\delta(d_\ell-d_{i}) \leq c_{\ell-1}+\delta(d_\ell-d_{\ell-1})$
		\label{alg:optconv:if1}}{
		$L_\ell \gets L_{\ell-1}$\;
  }\Else{
		$L_\ell \gets (\ell-1),L_{\ell-1}$\;
		$r \gets \ell$\;
		\While{top pair $(i,j)$ from $S$ satisfies 
			$c_{i}+\delta(d_{j}-d_{i}) > c_{\ell-1}+\delta(d_{j}-d_{\ell-1})$
      \label{alg:optconv:removebeg}}{
			$r \gets j$\;
			Remove $(i,j)$ from $S$
      \label{alg:optconv:removeend}\;
		}
		\If{$S$ is empty}{
			$S \gets (\ell-1,m+1)$\;}
		\Else{
			$(i,j) \gets$ top pair from $S$\;
			$v \gets$ least index with $r \leq v < j$ s.t. 
				$c_{\ell-1}+\delta(d_v-d_{\ell-1}) > c_{i}+\delta(d_v-d_{i})$
				\label{alg:optconv:findv}\;
			$S \gets (\ell-1,v),S$\;
    }
  }
	$c_\ell \gets c_{i}+\delta(d_\ell-d_{i})$ \qquad
		(where $(i,j)$ is top pair from $S$)\;
}
\Return{$L_{m+1}$}
\caption{Chunky Conversion Algorithm}
\label{alg:optconv}
\end{algorithm2e}

For an informal justification of correctness,
consider a single iteration through the main {\bf for} loop. At this
point, we have computed all optimal costs $c_1,c_2,\ldots,c_{\ell-1}$, and
the lists of gaps to achieve those costs $L_1,L_2,\ldots,L_{\ell-1}$. We
also have computed the stack $S$, indicating which of the gaps up to
index $\ell-2$ is optimal and when.

The {\bf while} loop on Step~\ref{alg:optconv:remold} removes all gaps
from the stack which are no longer relevant, either because their cost
is now beaten by a previous gap (when $j<\ell$), or because the size of
the resulting chunk would be greater than $k$ and therefore unnecessary
to consider.

If the condition of Step~\ref{alg:optconv:if1} is
true, then there is no index at which gap $(\ell-1)$ should be used,
so we discard it.

Otherwise, the gap at index $\ell-1$ is good at least some of the time, so
we proceed to the task of determining the largest gap index $v$ at which 
gap $(\ell-1)$ might still be useful. First, in 
Steps~\ref{alg:optconv:removebeg}--\ref{alg:optconv:removeend},
we repeatedly check whether gap $(\ell-1)$ always beats the gap at the top of
the stack $S$, and if so remove it. After this process, either no gaps remain
on the stack, or we have a range $r\leq v\leq j$ in which binary search can
be performed to determine $v$.

From the definitions, $d_{m+1}=\deg f+1$, and so the list of gaps
$L_{m+1}$ returned on the final step gives the optimal list of gaps to
include in $f \bmod x^{\deg f + 1}$, which is of course just $f$ itself.

\begin{thm}\label{thm:optconv}
Algorithm~\ref{alg:optconv} returns the optimal chunky representation for
multiplying $f$ by a dense size-$k$ chunk. The running time of the algorithm
is linear in the size of the input representation of $f$.
\end{thm}
\begin{pf}
Correctness follows from the discussions above.

For the complexity analysis, first note that
the maximal size of $S$, as well as
the number of saved values $a_i,b_i,d_i,s_i,L_i$, is $m$, the number of
gaps in $f$. Clearly $m$ is less than the number of nonzero terms in $f$, 
so this is bounded above by the
sparse or dense representation size.
If the lists $L_i$ are implemented as singly-linked lists, sharing nodes,
then the total extra storage for the algorithm is $O(m)$.

The total number of iterations of the two \textbf{while} loops corresponds to 
the number of gaps that are removed from the stack $S$ at any step. Since
at most one gap is pushed onto $S$ at each step, the total number of removals,
and hence the total cost of these \textbf{while} loops over all iterations,
is $O(m)$.

Now consider the cost of Step~\ref{alg:optconv:findv} at each iteration.
If the input is given in the sparse representation, we just perform a
binary search on the interval from $r$ to $j$, for a total cost of
$O(m\log m)$ over all iterations. Because $m$ is at most the number of
nonzero terms in $f$, $m\log m$ is bounded above
by the sparse representation size, so the theorem is satisfied for sparse
input.

When the input is given in the dense representation, we also use a
binary search for Step~\ref{alg:optconv:findv}, but we start with a
one-sided binary search, or ``galloping'' search, from either $r$ or
$j$, depending on which $v$ is closer to.
The cost of this search is at a single iteration is
$O(\log \min\{v-r,i_2-v\})$. Notice that the interval
$(r,j)$ in the stack is then effectively split at the index $v$, so
intuitively whenever more work is required through one iteration of this
step, the size of intervals is reduced, so future iterations should have
lower cost.

More precisely, 
a loose upper bound in the worst case of the total cost over all iterations is
$O(\sum_{i=1}^u 2^i \cdot (u-i+1))$, where 
$u = \lceil \log_2 m \rceil$. This is less than $2^{u+2}$, which is $O(m)$,
giving linear cost in the size of the
dense representation.
\end{pf}

\subsection{Determining the optimal chunk size}

All that remains is to compute the optimal chunk size $k$ that will be
used in the conversion algorithm from the previous section. 
This is accomplished by finding the value of $k$ that minimizes the cost
of multiplying two polynomials $f,g\in\R[x]$, under the restriction that
every chunk of $f$ and of $g$ has size $k$.

If $f$ is written in the chunky representation as in \eqref{eqn:chunkyrep},
there are many possible choices for the number of chunks $t$, depending on
how large the chunks are. So define $t(k)$ to be the least number of chunks
if each chunk has size at most $k$, i.e. $\deg f_i<k$ for $1\leq i \leq t(k)$.
Similarly define $s(k)$ for $g\in\R[x]$ written as in
\eqref{eqn:chunkyg}.

Therefore, from the cost of multiplication in Theorem~\ref{thm:chunkycost},
in this part we want to compute the value of $k$ that minimizes
\begin{equation} \label{eqn:kcost}
t(k) \cdot s(k) \cdot k \cdot \delta(k).
\end{equation}

Say $\deg f = n$. After $O(n)$ preprocessing work (making pointers to the
beginning and end of each ``gap''), $t(k)$ could be computed using
$O(n/k)$ word operations, for any value $k$. This leads to one possible
approach to computing the value of $k$ that minimizes \eqref{eqn:kcost}
above: simply compute \eqref{eqn:kcost} for each possible
$k=1,2,\ldots,\max\{\deg f, \deg g\}$.
This na\"ive approach is too costly for our purposes, but underlies the basic
idea of our algorithm.

Rather than explicitly computing each $t(k)$ and $s(k)$, we essentially
maintain chunky representations of $f$ and $g$ with all chunks having
size less than $k$, starting with $k=1$. As $k$
increases, we count the number of chunks in each representation, which
gives a tight approximation to the actual values of $t(k)$
and $f(k)$, 
while achieving
linear complexity in the size of
either the sparse or dense representation.

To facilitate the ``update'' step, a minimum priority queue $Q$ 
(whose specific
implementation depends on the input polynomial representation)
is maintained containing all gaps in the current chunky representations
of $f$ and $g$. For each gap, the key value (on which the priority queue
is ordered) is the size of the chunk that would result from merging the
two chunks adjacent to the gap into a single chunk.

So for example, if we write $f$ in the chunky representation as
\[f = (4 + 0x + 5x^2)\cdot x^{12} + (7 + 6x + 0x^2 + 0x^3 + 8x^4)\cdot x^{50},\]
then the single gap in $f$ will have key value $3+35+5=43$, 
More precisely, if $f$ is written
as in \eqref{eqn:chunkyrep}, then the $i^{\rm th}$ gap has key value
\begin{equation}\label{eqn:gapkey}\deg f_{i+1}+e_{i+1}-e_i.+1\end{equation}

Each gap in the priority queue also contains pointers to the two 
(or fewer) neighboring 
gaps in the current chunky representation. 
Removing a gap from the queue corresponds to 
merging the two chunks adjacent to that
gap, so we will need to update (by increasing) the key values of any
neighboring gaps accordingly.

At each iteration through the main loop in the algorithm, the
smallest key value in the priority queue is examined, and $k$ is increased
to this
value. Then gaps with key value $k$ are repeatedly removed from the queue
until no more remain. This means that each remaining gap, if removed, would
result in a chunk of size strictly greater than $k$. Finally, we compute
$\delta(k)$ and an approximation of \eqref{eqn:kcost}.

Since the purpose here is only to compute an optimal chunk size
$k$, and not actually to compute chunky representations of $f$ and $g$, we
do not have to maintain chunky representations of the polynomials
as the algorithm proceeds, but merely counters for the number of chunks in
each one. Algorithm~\ref{alg:findk} gives the details of this computation.

  \begin{algorithm2e}[htb]
\DontPrintSemicolon
\KwIn{$f,g\in\R[x]$}
\KwOut{$k\in\NN$ that minimizes $t(k)\cdot s(k)\cdot k\cdot \delta(k)$}
$Q_f,Q_g \gets$ minimum priority queues initialized with all gaps in $f$
and $g$, respectively\;
$k,k_{\rm min} \gets 1; \quad c_{\rm min} \gets t_f t_g$\;
\While{$Q_f$ and $Q_g$ are not both empty}{
	$k \gets$ smallest key value from $Q_f$ or $Q_g$
		\label{alg:findk:smallestk}\;
	\While{$Q_f$ has an element with key value $\leq k$\label{alg:findk:whilef}}{
		Remove a $k$-valued gap from $Q_f$ and update neighbors\;
	}
	\While{$Q_g$ has an element with key value $\leq k$}{
		Remove a $k$-valued gap from $Q_g$ and update neighbors\;
	}
	$c_{\rm current} \gets (|Q_f|+1)\cdot(|Q_g|+1)\cdot k\cdot\delta(k)$\;
	\If{$c_{\rm current} < c_{\rm min}$
		\label{alg:findk:testmin}}{
		$k_{\rm min} \gets k; \quad 
			c_{\rm min} \gets c_{\rm current}$\;
	}
}
\Return{$k_{\rm min}$}
\caption{Optimal Chunk Size Computation}
\label{alg:findk}
\end{algorithm2e}

All that remains is the specification of the data structures used to
implement the priority queues $Q_f$ and $Q_g$. If the input polynomials
are in the sparse representation, we simply use standard binary heaps,
which give logarithmic cost for each removal and update.
Because the exponents in this case are multi-precision integers, we
might imagine encountering chunk sizes that are larger than the largest
word-sized integer. But as discussed previously, such a chunk size
would be meaningless since a dense polynomial with that size
cannot be represented in memory. So our priority queues may discard any
gaps whose key value is larger than word-sized. This guarantees all keys
in the queues are word-size integers, which is necessary for the
complexity analysis later.

If the input polynomials are dense, we need a structure which
can perform removals and updates in constant time, using
$O(\deg f + \deg g)$ time and space. For $Q_f$, we use an
array with length $\deg f$ of (possibly empty) linked lists, where the
list at index $i$ in the array contains all elements in the queue with
key $i$. (An array of this length is sufficient because each key value
in $Q_f$ is at least 2 and at most $1+\deg f$.) We use the same data
structure for $Q_g$, and this clearly gives constant time for each
remove and update operation.

To find the smallest key value in either queue at each
iteration through
Step~\ref{alg:findk:smallestk}, we simply start at the beginning of the
array and search forward in each position until a non-empty list is
found. Because each queue element update only results in the key values
\emph{increasing}, we can start the search at each iteration at the
point where the previous search ended. Hence the total cost of
Step~\ref{alg:findk:smallestk} for all iterations is 
$O(\deg f+\deg g)$.

The following lemma proves that our approximations of $t(k)$ and
$s(k)$ are reasonably tight, and will be crucial in proving the
correctness of the algorithm.

\begin{lem}\label{lem:approxc}
At any iteration through Step~\ref{alg:findk:testmin} in
Algorithm~\ref{alg:findk}, $|Q_f| < 2t(k)$ and 
$|Q_g| < 2s(k)$.
\end{lem}
\begin{pf}
	First consider $f$.
	There are two chunky representations with each chunk of degree less
	than $k$ to consider: the optimal having $t(k)$ chunks
	and the one implicitly computed by Algorithm~\ref{alg:findk}
  with $|Q_f|+1$ chunks. Call these
	$\bar{f}$ and $\hat{f}$, respectively.
  
  We claim that any single chunk of the optimal $\bar{f}$ contains at most
  three constant terms of chunks in the implicitly-computed $\hat{f}$.
  If this were not so, then two chunks in $\hat{f}$ could be combined to result
  in a single chunk with degree less than $k$. But this is impossible, since
  all such pairs of chunks would already have been merged after the completion
  of Step~\ref{alg:findk:whilef}.

	Therefore every chunk in $\bar{f}$ contains at most two
	constant terms of distinct chunks in $\hat{f}$. Since each constant 
	term of a chunk is required to be nonzero, 
	the number of chunks in $\hat{f}$ is at most
	twice the number in $\bar{f}$. Hence
	$|Q_f| +1 \leq 2t(k)$. An identical argument for $g$ gives the stated
	result.
\end{pf}

Now we are ready for the main result of this subsection.

\begin{thm}\label{thm:optk}
Algorithm~\ref{alg:findk} computes a chunk size $k$ such that
$t(k)\cdot s(k)\cdot k\cdot \delta(k)$ is at most 4 times the
minimum value. The worst-case cost of the algorithm
is linear in the size of the input representations.
\end{thm}
\begin{pf}
	If $k$ is the value returned from the algorithm and $k^*$ is the
	value which actually minimizes \eqref{eqn:kcost}, the worst that can
	happen is that the algorithm computes the actual value of
	$c_f(k)\,c_g(k)\,k\,\delta(k)$, but overestimates the value of
	$c_f(k^*)\,c_g(k^*)\,k^*\,\delta(k^*)$. This overestimation can only
	occur in $c_f(k^*)$ and $c_g(k^*)$, and each of those by only a
	factor of 2 from Lemma~\ref{lem:approxc}. 
  So the first statement of the theorem holds.

  Write $c$ for the total number of nonzero terms in $f$ and $g$.
  The initial sizes of the queues $Q_f$ and $Q_g$ is $O(c)$.
	Since gaps are only removed from the queues (after they are initialized),
  the total cost of all queue operations is bounded above by $O(c)$,
  which in turn is bounded above by the sparse
  and dense sizes of the input polynomials.

	If the input is sparse and we use a binary heap, the cost of each
	queue operation is $O(\log c)$, for a total cost of $O(c\log c)$,
	which is a lower bound on the size of the sparse representations. If
	the input is in the dense representation, then each queue operation
	has constant cost. Since $c\in O(\deg f+\deg g)$, the total
	cost linear in the size of
	the dense representation.
\end{pf}

\subsection{Chunky Multiplication Overview}

Now we are ready to examine the whole process of
chunky polynomial conversion and
multiplication. First we need the following easy corollary of
Theorem~\ref{thm:optconv}.

\begin{cor}\label{cor:optconv}
Let $f\in\R[x]$, $k\in\NN$,
and $\hat{f}$ be any chunky representation of $f$ where all chunks have
degree at least $k$, and $\bar{f}$ be the representation returned by
Algorithm~\ref{alg:optconv} on input $k$.
The cost of multiplying $\bar{f}$ by a single chunk of size $\ell<k$ is
then less than the cost of multiplying $\hat{f}$ by the same chunk.
\end{cor}
\begin{pf}
Consider the result of
Algorithm~\ref{alg:optconv} on input $\ell$. We know from
Theorem~\ref{thm:optconv} that this gives the optimal chunky
representation for multiplication of $f$ with a size-$\ell$ chunk.
But the only difference in the algorithm on input $\ell$ and input $k$ 
is that more pairs are removed at each iteration on
Step~\ref{alg:optconv:remold} on input $\ell$.

This means that every gap included in the representation $\bar{f}$ is
also included in the optimal representation. We also know that all
chunks in $\bar{f}$ have degree less than $k$, so that $\hat{f}$ must
have fewer gaps that are in the optimal representation than $\bar{f}$.
It follows that multiplication of a size-$\ell$ chunk by $\bar{f}$ is
more efficient than multiplication by $\hat{f}$.
\end{pf}

To review, the entire process to multiply $f,g\in\R[x]$ using the chunky
representation is as follows:

\begin{enumerate}
\item Compute $k$ from Algorithm~\ref{alg:findk}.
\item Compute chunky representations of $f$ and $g$
	using Algorithm~\ref{alg:optconv} with input $k$.
\item Multiply the two chunky representations using
Algorithm~\ref{alg:chunkymul}.
\item Convert the chunky result back to the original representation.
\end{enumerate}

Because each step is optimal (or within a constant bound of the
optimal), we expect this approach to yield the most efficient chunky
multiplication of $f$ and $g$. In any case, we know it will be at least
as efficient as the standard sparse or dense algorithm.

\begin{thm}
Computing the product of $f,g\in\R[x]$ never uses more ring operations
than either the standard sparse or dense polynomial multiplication algorithms.
\end{thm}
\begin{pf}
In Algorithm~\ref{alg:findk}, the values of
$t(k)\cdot s(k)\cdot k\cdot \delta(k)$ for $k=1$ and $k=\min\{\deg
f, \deg g\}$ correspond to the costs of the standard sparse and dense
algorithms, respectively. Furthermore, it is easy to see that these
values are never overestimated, meaning that the $k$ returned from the
algorithm which minimizes this formula gives a cost which is not greater
than the cost of either standard algorithm.

Now call $\hat{f}$ and $\hat{g}$ the implicit representations from
Algorithm~\ref{alg:findk}, and $\bar{f}$ and $\bar{g}$ the
representations returned from Algorithm~\ref{alg:optconv} on input $k$.
We know that the multiplication of $\hat{f}$ by $\hat{g}$ is more
efficient than either standard algorithm from above. Since every chunk
in $\hat{g}$ has size $k$, multiplying $\bar{f}$ by $\hat{g}$ will have
an even lower cost, from Theorem~\ref{thm:optconv}. Finally, since every
chunk in $\bar{f}$ has size at most $k$, Corollary~\ref{cor:optconv}
tells us that the cost is further reduced by multiplying $\bar{f}$ by
$\bar{g}$.

The proof is complete from the fact that conversion back to either
original representation takes linear time in the size of the output.
\end{pf}

\section{Equal-Spaced Polynomials}
\label{sec:eqspace}

Next we consider an adaptive representation which is in some sense orthogonal
to the chunky representation. This representation will be useful when the
coefficients of the polynomial are not grouped together into dense chunks,
but rather when they are spaced evenly apart.

Let $f \in \R[x]$ with degree $n$, and 
suppose the exponents of $f$ are all divisible by some integer $k$. 
Then we can write
$f = a_0 + a_1x^k + a_2x^{2k} + \cdots$. So by letting
$f_D = a_0 + a_1 x + a_2 x^2 + \cdots$, we have
$f = f_D\circ x^k$ (where the symbol $\circ$ indicates functional
composition).

One motivating example suggested by Michael Monagan is that of homogeneous
polynomials. Recall that a multivariate polynomial $h\in\R[x_1,\ldots,x_n]$
is \emph{homogeneous of degree $d$} if every nonzero term of $h$ has
total degree $d$. It is well-known that the number of variables in a homogeneous
polynomial can be effectively reduced by one by writing
$y_i = x_i/x_n$ for $1\leq i < n$ and 
$h = x_n^{\phantom{n}d} \cdot \hat{h}$, for
$\hat{h}\in\R[y_1,\ldots,y_{n-1}]$ an $(n-1)$-variate polynomial with
max-degree $d$. This leads to efficient schemes for homogeneous polynomial
arithmetic.

But this is only possible if (1) the user realizes this structure in their
polynomials, and (2) every polynomial used is homogeneous. Otherwise, a more
generic approach will be used,
such as the Kronecker substitution mentioned in the
introduction. Choosing some integer $\ell>d$, we evaluate
$h(y,y^\ell,y^{\ell^2},\ldots,y^{\ell^{n-1}})$, 
and then perform univariate arithmetic
over $\R[y]$. But if $h$ is homogeneous, a special structure arises:
every exponent of $y$ is of the form $d + i(\ell-1)$ for some
integer $i\geq 0$. Therefore we can write 
$h(y,\ldots,y^{\ell^{n-1}}) = (\bar{h} \circ y^{\ell-1})\cdot y^d$, for
some $\bar{h}\in\R[y]$ with much smaller degree. The algorithms presented
in this section will automatically recognize this structure and perform
the corresponding optimization to arithmetic.

The key idea is \emph{equal-spaced representation},
which corresponds to writing a polynomial $f\in\R[x]$ as
\begin{equation}
\label{eqn:eqspace}
f = (f_D\circ x^k)\cdot x^d + f_S,
\end{equation}
with 
$k,d \in \mathbb{N}$, $f_D\in\R[x]$ dense with degree less than $n/k-d$, 
and $f_S\in\R[x]$ sparse with degree less than $n$. The polynomial $f_S$ is a
``noise'' polynomial which contains the comparatively few terms in $f$
whose exponents are not of the form $ik+d$ for some $i\geq 0$.

Unfortunately,
converting a sparse polynomial to the best equal-spaced representation
seems to be difficult. To see why this is the case, consider the much
simpler problem of verifying that a sparse polynomial $f$ can be written
as $(f_D\circ x^k)\cdot x^d$. For each exponent $e_i$ of a nonzero term
in $f$, this means confirming that $e_i \equiv d \bmod k$. But the cost
of computing each $e_i \bmod k$ is roughly $O(\sum (\log e_i) \delta(\log k))$,
which is a factor of $\delta(\log k)$ greater than the size of the
input. Since $k$ could be as large as the exponents, we see that even
verifying a proposed $k$ and $d$ takes too much time for the conversion
step. Surely computing such a $k$ and $d$ would be even more costly!

Therefore, for this subsection,
we will always assume that the input polynomials are given in the dense
representation. In Section~\ref{sec:combine}, we will see how by combining
with the chunky representation, we effectively handle equal-spaced sparse
polynomials without ever having to convert a sparse polynomial directly to
the equal-spaced representation.

\subsection{Multiplication in the equal-spaced representation}
Let $g\in\R[x]$ with degree less than $m$ and write
$g=(g_D\circ x^\ell)\cdot x^e + g_S$ as in \eqref{eqn:eqspace}.
To compute $f\cdot g$, simply 
sum up the four pairwise products of terms. All these except for the
product $(f_D\circ x^k)\cdot (g_D \circ x^\ell)$ 
are performed using standard sparse
multiplication methods.

Notice that if $k=\ell$, then $(f_D\circ x^k)\cdot (g_D \circ x^\ell)$
is simply $(f_D\cdot g_D)\circ x^k$, and hence is efficiently computed using
dense multiplication. However, if $k$ and $\ell$
are relatively prime, then
almost any term in the product can be nonzero. 

This indicates that the gcd of $k$ and $\ell$ is very significant. Write
$r$ and $s$ for the greatest common divisor and least common multiple of
$k$ and $\ell$, respectively. 
To multiply $(f_D\circ x^k)$ by $(g_D\circ x^\ell)$, we perform
a transformation similar to the process of finding common denominators in
the addition of fractions. First
split $f_D\circ x^k$ into $s/k$ (or $\ell/r$) polynomials, each with degree less
than $n/s$ and right composition factor $x^s$, as follows:
\[f_D\circ x^k = (f_0\circ x^s) + (f_1 \circ x^s) \cdot x^k + 
	(f_2 \circ x^s) \cdot x^{2k} \cdots
	+ (f_{s/k-1}\circ x^s) \cdot x^{s-k} \]

Similarly split $g_D\circ x^\ell$ into $s/\ell$ polynomials 
$g_0,g_1,\ldots,g_{s/\ell-1}$ with degrees less than $m/s$ and right
composition factor $x^s$.
Then compute all pairwise products $f_i\cdot g_j$,
and combine them appropriately to compute the total sum (which will be
equal-spaced with right composition factor $x^r$).

Algorithm~\ref{alg:eqmul} gives the details of this method.

\begin{algorithm2e}[htb]
\DontPrintSemicolon
\KwIn{$f = (f_D\circ x^k)\cdot x^d + f_S, 
	\quad g = (g_D\circ x^\ell)\cdot x^e + g_S$,\\
	with $f_D = a_0 + a_1x + a_2x^2 + \cdots, 
	\quad g_D = b_0 + b_1x + b_2x^2 + \cdots$}
\KwOut{The product $f \cdot g$}
$r \gets \gcd(k,\ell),\quad s \gets \lcm(k,\ell)$\;
\For{ $i = 0,1,\ldots,s/k - 1$ \label{alg:eqmul:1}}{
$f_i \gets a_i + a_{s+i}x + a_{2s+i} x^2 + \cdots$\;
}
\For{ $i = 0,1,\ldots,s/\ell - 1$ }{
$g_i \gets b_i + b_{s+i}x + b_{2s+i} x^2 + \cdots$\;
\label{alg:eqmul:2}}
$h_D \gets 0$
\label{init-h1}\;
\For{ $i = 0,1,\ldots,s/k-1$ }{
\For{ $j = 0,1,\ldots,s/\ell-1$ }{
Compute $f_i \cdot g_j$ by dense multiplication
\label{alg:eqmul:mul}\;
$h_D \gets h_D + ((f_i \cdot g_j)\circ x^s)\cdot x^{ik+j\ell}$
\label{alg:eqmul:add1}\;
}
}
Compute $(f_D\circ x^k)\cdot g_S, (g_D\circ x^\ell)\cdot f_S,$
	and $f_S \cdot g_S$ by sparse multiplication
\label{alg:eqmul:init}\;
\Return{$h_D\cdot x^{e+d} + (f_D\circ x^k)\cdot g_S\cdot x^d +
	(g_D\circ x^\ell)\cdot f_S\cdot x^e + f_S \cdot g_S$
\label{alg:eqmul:add2}}
\caption{Equal Spaced Multiplication}
\label{alg:eqmul}
\end{algorithm2e}

As with chunky multiplication, this final product is easily converted to
the standard dense representation in linear time. The
following theorem gives the complexity analysis for equal-spaced multiplication.

\begin{thm}
\label{thm:eqmul}
Let $f,g$ be as above such that $n > m$,
and write $t_f,t_g$ for the number of 
nonzero terms in $f_S$ and $g_S$, respectively. 
Then Algorithm~\ref{alg:eqmul} correctly computes the product $f\cdot g$
using
\[O\left( (n/r) \cdot \delta(m/s) + n t_g/k + mt_f/\ell + t_ft_g \right)\]
ring operations.
\end{thm}
\begin{pf}
Correctness follows from the preceding discussion.

The polynomials $f_D$ and $g_D$ have at most $n/k$ and $m/\ell$
nonzero terms,
respectively. So the cost of computing the three products 
in Step~\ref{alg:eqmul:init} by using standard sparse multiplication is
$O( n t_g/k + mt_f/\ell+ t_ft_g)$ ring operations,
giving the last three terms in the complexity measure.

The initialization in Steps~\ref{alg:eqmul:1}--\ref{alg:eqmul:2} and the
additions in Steps~\ref{alg:eqmul:add1} and \ref{alg:eqmul:add2} all
have cost bounded by $O(n/r)$, and hence do not dominate the complexity.

All that remains is the cost of computing each product $f_i \cdot g_j$
by dense multiplication on Step~\ref{alg:eqmul:mul}. From the discussion
above, $\deg f_i < n/s$ and $\deg g_j < m/s$, for each $i$ and $j$.
Since $n>m$, $(n/s)>(m/s)$, and therefore this product can be computed
using $O((n/s)\delta(m/s))$ ring operations. The number of iterations
through Step~\ref{alg:eqmul:mul} is exactly $(s/k)(s/\ell)$. But
$s/\ell=k/r$, so the number of iterations is just $s/r$. Hence the total
cost for this step is $O((n/r)\delta(m/s))$,
which gives the first term in the complexity measure.
\end{pf}

It is worth noting that no additions of ring elements are actually
performed through each iteration of Step~\ref{alg:eqmul:add1}. The proof
is as follows. If any additions were performed, we would have
$$i_1k+j_1\ell \equiv i_2k+j_2\ell \mod s$$ for distinct pairs
$(i_1,j_1)$ and $(i_2,j_2)$. Without loss of generality, assume 
$i_1\neq i_2$, and write
$$(i_1k+j_1\ell) - (i_2k+j_2\ell) = qs$$ 
for some $q\in\ZZ$. Rearranging gives
$$(i_1-i_2)k = (j_2-j_1)\ell + qs.$$
Because $\ell|s$, the left hand side is
a multiple of both $k$ and $\ell$, and therefore by definition must be a
multiple of $s$, their lcm. Since $0 \leq i_1,i_2 < s/k$,
$|i_1-i_2|<s/k$, and therefore $|(i_1-i_2)k|<s$. The only multiple of
$s$ with this property is of course 0, and since $k\neq 0$ this means
that $i_1=i_2$, a contradiction.

The following theorem compares the cost of equal-spaced multiplication to
standard dense multiplication, and will be used to guide the approach to
conversion below.

\begin{thm}\label{thm:eqspnoise}
Let $f,g,m,n,t_f,t_g$ be as before. Algorithm~\ref{alg:eqmul} does not
use asymptotically more
ring operations than standard dense multiplication to compute the
product of $f$ and $g$ as long as $t_f \in O(\delta(n))$ and 
$t_g \in O(\delta(m))$.
\end{thm}
\begin{pf}
	Assuming again that $n>m$, the cost of standard dense multiplication
	is $O(n\delta(m))$ ring operations, which is the same as
	$O(n\delta(m)+m\delta(n))$.

	Using the previous theorem, the number of ring operations used by
	Algorithm~\ref{alg:eqmul} is 
	\[O\left((n/r) \delta(m/s) + n \delta(m)/k + m\delta(n)/\ell +
	\delta(n)\delta(m) \right).\]

	Because all of $k,\ell,r,s$ are at least 1, and since $\delta(n)<n$,
	every term in this complexity measure is bounded by
	$n\delta(m)+m\delta(n)$. The stated result follows.
\end{pf}

\subsection{Converting to equal-spaced}
The only question when converting a polynomial $f$
to the equal-spaced representation
is how large we should allow $t_S$ (the number of nonzero terms in of $f_S$)
to be. From Theorem~\ref{thm:eqspnoise} above, clearly we need 
$t_S \in \delta(\deg f)$, but we can see from the proof of the theorem that
having this bound be tight will often give performance that is equal to the
standard dense method (not worse, but not better either).

Let $t$ be the number of nonzero terms in $f$.
Since the goal of any adaptive method is to in fact be faster than the
standard algorithms, we use the lower bound of $\delta(n) \in \Omega(\log n)$
and $t \leq \deg f + 1$
and require that $t_S<\log_2 t$.

As usual, let $f\in\R[x]$ with degree less than $n$ and write
\[f = a_1 x^{e_1} + a_2 x^{e_2} + \cdots + a_t x^{e_t},\]
with each $a_i \in \R\setminus \{0\}$. The reader will recall that this
corresponds to the sparse representation of $f$, but keep in mind that
we are assuming $f$ is given in the dense representation; 
$f$ is written this way only for notational convenience.

The conversion problem is then to find the largest possible value of $k$
such that all but at most $\log_2 t$ of the exponents $e_j$ can be
written as $ki+d$, for any nonnegative integer $i$ and a fixed integer $d$.
Our approach to computing $k$ and $d$ will be simply to check each
possible value of $k$, in decreasing order. To make this efficient, we
need a bound on the size of $k$.

\begin{lem}\label{lem:kbound}
Let $n\in\NN$ and
$e_1,\ldots,e_t$ be distinct integers in the range $[0,n]$.
If at least $t-\log_2 t$ of the integers $e_i$ are congruent to the same
value modulo $k$, for some $k\in\NN$, then
\[k \leq \frac{n}{t-2\log_2 t - 1}.\]
\end{lem}
\begin{pf}
Without loss of generality, order the $e_i$'s so that 
$0\leq e_1 < e_2 < \cdots < e_t \leq n$.
Now consider the telescoping sum
$(e_2-e_1) + (e_3-e_2) + \cdots + (e_t-e_{t-1})$.
Every term in the sum is at least 1, and the total is $e_t-e_1$, which
is at most $n$.

Let $S \subseteq \{e_1,\ldots,e_t\}$ be the set of at most $\log_2 t$
integers not congruent to the others modulo $k$. 
Then for any $e_i,e_j \notin S$, $e_i\equiv e_j\bmod k$. Therefore
$k|(e_j-e_i)$. If $j>i$, this means that $e_j-e_i\geq k$.

Returning to the telescoping sum above, each $e_j\in S$ is in at most
two of the sum terms $e_i-e_{i-1}$. So all but at most $2\log_2 t$ of
the terms are at least $k$. Since there are exactly $t-1$ terms, and the
total sum is at most $n$, we conclude that
$(t-2\log_2 t-1)\cdot k \leq n$. The stated result follows.
\end{pf}

We now employ this lemma to develop an algorithm to determine the best
values of $k$ and $d$, given a dense polynomial $f$. Starting from the
largest possible value from the bound, for each candidate value $k$, we
compute each $e_i \bmod k$, and find the majority element --- that is,
a common modular image of more than half of the exponents.

To compute the majority element, we use a now well-known approach first
credited to \citet{BoyMoo81} and \citet{FisSal82}. Intuitively, pairs of
different elements are repeatedly removed until only one element
remains. If there is a majority element, this remaining element is it;
only one extra pass through the elements is required to check whether 
this is the case. In practice, this is accomplished without actually
modifying the list.

\begin{algorithm2e}[htb]
\DontPrintSemicolon
\KwIn{Exponents $e_1,e_2,\ldots,e_t \in \NN$ and $n\in\NN$ such
that $0\leq e_1 < e_2 < \cdots < e_t = n$}
\KwOut{$k,d \in \NN$ and $S \subseteq \{e_1,\ldots,e_t\}$ such that
$e_i \equiv d \bmod k$
for all exponents $e_i$ not in $S$, and
$|S| \leq \log_2 t$.}
\lIf{$t < 32$}{$k \gets n$ \label{alg:eqspconv:setk1}}\;
\lElse{$k \gets \lfloor n/(t-1-2\log_2 t) \rfloor$ 
	\label{alg:eqspconv:setk2}
}\;
\While{$k \geq 2$}{
	$d \gets e_1 \bmod k; \quad j\gets 1$\;
	\For{$i=2,3,\ldots,t$}{
		\lIf{$e_i \equiv d \bmod k$}{
			$j \gets j+1$
    }\;
		\lElseIf{$j>0$}{
			$j \gets j-1$
    }\;
		\lElse{
			$d \gets e_i \bmod k; \quad j \gets 1$
		}\;
	}
	$S \gets \{e_i : e_i \not\equiv d \bmod k\}$
		\label{alg:eqspconv:S}\;
	\lIf{$|S| \leq \log_2 t$}{
		\Return{$k,\,d,\,S$}
	}\;
	$k \gets k-1$\;
}
\Return{$1,\,0,\,\emptyset$}
\caption{Equal Spaced Conversion}
\label{alg:eqspconv}
\end{algorithm2e}

Given $k,d,S$ from the algorithm, in one more pass through
the input polynomial,
$f_D$ and $f_S$ are constructed such that $f=(f_D\circ x^k)\cdot x^d + f_S$. 
After performing separate
conversions for two polynomials $f,g\in\R[x]$, they are
multiplied using Algorithm~\ref{alg:eqmul}.

The following theorem proves correctness when $t>4$. If $t\leq 4$, we
can always trivially set $k=e_t-e_1$ and $d=e_1\bmod k$ to satisfy the stated
conditions.

\begin{thm}\label{thm:eqspconv}
Given integers $e_1,\ldots,e_t$ and $n$, with $t > 4$, 
Algorithm~\ref{alg:eqspconv}
computes the largest integer $k$ such that at least $t-\log_2 t$ of the
integers $e_i$ are congruent modulo $k$, and uses $O(n)$ word
operations.
\end{thm}
\begin{pf}
In a single iteration through the \textbf{while} loop, we compute the
majority element of the set $\{e_i \bmod k : i=1,2,\ldots,t\}$, if there
is one. 
Because
$t>4$, $\log_2 t < t/2$. Therefore any element which occurs at least
$t-\log_2 t$ times in a $t$-element set is a majority element, which
proves that any $k$ returned by the algorithm is such that at least
$t-\log_2 t$ of the integers $e_i$ are congruent modulo $k$.

From
Lemma~\ref{lem:kbound}, we know that the initial value of $k$ on
Step~\ref{alg:eqspconv:setk1} or \ref{alg:eqspconv:setk2} is greater
than the optimal $k$ value. Since we start at this value and decrement
to 1, the largest $k$ satisfying the stated conditions is returned.

For the complexity analysis, first consider the cost of a single iteration
through the main \textbf{while} loop. Since each integer
$e_i$ is word-sized, computing each $e_i \bmod k$ has constant cost, 
and this happens $O(t)$ times in each iteration.

If $t<32$, each of the $O(n)$ iterations has constant cost, for total cost
$O(n)$.

Otherwise, we start with $k=\lfloor n/(t-1-2\log_2 t) \rfloor$ and
decrement. Because $t\geq 32$, $t/2 > 1 + 2\log_2 t$. Therefore 
$(t-1-2\log_2 t) > t/2$, so the initial value of $k$ is less than
$2n/t$. This gives an upper bound on the number of iterations through
the \textbf{while} loop, and so the total cost is $O(n)$ word
operations, as required.
\end{pf}

Algorithm~\ref{alg:eqspconv} can be implemented using only 
$O(t)$ space for the storage of the exponents $e_1,\ldots,e_t$,
which is linear in the size of the output, plus the space
required for the returned set $S$.

\section{Chunks with Equal Spacing}
\label{sec:combine}
The next question is whether the ideas of chunky and equal-spaced
polynomial multiplication can be effectively combined into a single
algorithm. As before, we seek an \emph{adaptive} combination of previous
algorithms, so that the combination is never asymptotically worse than
either original idea.

An obvious approach would be to first
perform chunky polynomial conversion, and then equal-spaced conversion on
each of the dense chunks. 
Unfortunately, this would be asymptotically 
less efficient than equal-spaced multiplication
alone in a family of instances, and therefore is not acceptable as a
proper adaptive algorithm.

The algorithm presented here
does in fact perform chunky conversion first, but instead of
performing equal-spaced conversion on each dense chunk independently,
Algorithm~\ref{alg:eqspconv} is run simultaneously on all chunks
in order to determine, for each polynomial, a single
spacing parameter $k$ that will be used for every chunk.

Let $f = f_1x^{e_1} + f_2x^{e_2} + \cdots + f_tx^{e_t}$ in the optimal
chunky representation for multiplication by another polynomial $g$. 
We first compute the smallest bound on the spacing parameter
$k$ for any of the chunks $f_i$,
using Lemma~\ref{lem:kbound}. Starting with this value, we execute the
\textbf{while} loop of Algorithm~\ref{alg:eqspconv} for each polynomial
$f_i$, stopping at the largest value of $k$ 
such that the total size of all sets $S$ on Step~\ref{alg:eqspconv:S}
for all chunks $f_i$ is at most $\log_2 t_f$, where $t_f$ is the 
total number
of nonzero terms in $f$.

The polynomial $f$ can then be rewritten (recycling the variables
$f_i$ and $e_i$) as
\[f= (f_1\circ x^k)\cdot x^{e_1} + (f_2\circ x^k)\cdot x^{e_2} + \cdots
	+ (f_t\circ x^k)\cdot x^{e_t} + f_S,\]
where $f_S$ is in the sparse representation and has $O(\log t_f)$
nonzero terms.

Let $k^*$ be the value returned from Algorithm~\ref{alg:eqspconv} on input
of the entire polynomial $f$. Using $k^*$ instead of $k$, $f$ could
still be written as above with $f_S$ having at most $\log_2 t_f$ terms.
Therefore the value of $k$ computed in this way is always greater than
or equal to $k^*$ if the initial bounds are correct. This will be the
case except when every chunk $f_i$ has few nonzero terms (and therefore
$t$ is close to $t_f$). However, this reduces to the problem of
converting a sparse polynomial to the equal-spaced representation, which
seems to be intractable, as discussed above. So our cost analysis will
be predicated on the assumption that the computed value of $k$ is never
smaller than $k^*$.

We perform the same equal-spaced conversion for $g$, and then use
Algorithm~\ref{alg:chunkymul} to compute the product $f\cdot g$,
with the difference that each product $f_i \cdot g_j$ is computed
by Algorithm~\ref{alg:eqmul} rather than standard dense
multiplication. As with equal-spaced multiplication, the products
involving $f_S$ or $g_S$ are performed using standard sparse
multiplication.

\begin{thm}\label{thm:eqspchunks}
The algorithm described above to multiply polynomials with equal-spaced
chunks never uses more ring operations than either chunky or
equal-spaced multiplication, provided that the computed ``spacing
parameters'' $k$ and $\ell$ are not smaller than the values returned
from Algorithm~\ref{alg:eqspconv}.
\end{thm}
\begin{pf}
Let $n,m$ be the degrees of $f,g$ respectively and 
write $t_f,t_g$ for the number of
nonzero terms in $f,g$ respectively. The sparse multiplications
involving $f_S$ and $g_S$ use a total of
$t_g \log t_f + t_f \log t_g + (\log t_f)(\log t_g)$ ring operations. 
Both the chunky or equal-spaced multiplication algorithms always
require $O(t_g\delta(t_f)+t_f\delta(t_g))$ ring operations in the best
case, and since $\delta(n)\in\Omega(\log n)$, the cost of these sparse
multiplications is never more than the cost of the standard chunky or
equal-spaced method.

The remaining computation is that to compute each product $f_i \cdot
g_j$ using equal-spaced multiplication. Write $k$ and $\ell$ for the
powers of $x$ in the right composition factors of $f$ and $g$
respectively. Theorem~\ref{thm:eqmul} tells us that the cost of
computing each of these products by equal-spaced multiplication is never
more than computing them by standard dense multiplication, since $k$ and
$\ell$ are both at least 1. Therefore the combined approach is never
more costly than just performing chunky multiplication.

To compare with the cost of equal-spaced multiplication, 
assume that $k$ and $\ell$ are the actual values returned by
Algorithm~\ref{alg:eqspconv} on input $f$ and $g$. This is the worst
case, since we have assumed that $k$ and $\ell$ are never smaller than
the values from Algorithm~\ref{alg:eqspconv}.

Now consider the
cost of multiplication by a single equal-spaced chunk of $g$.
This is the same as assuming $g$ consists of only one equal-spaced
chunk.
Write $d_i = \deg f_i$ for each equal-spaced chunk of $f$, and $r,s$ for
the gcd and lcm of $k$ and $\ell$, respectively. 
If $m > n$, then of course $m$ is larger than each $d_i$, so
multiplication using the combined method will use 
$O((m/r)\sum \delta(d_i/s))$ ring operations, compared to
$O((m/r) \delta(n/s))$ for the standard equal-spaced algorithm, by
Theorem~\ref{thm:eqmul}. 

Now recall the cost equation \eqref{eqn:kcost} 
used for Algorithm~\ref{alg:findk}:
\[c_f(b) \cdot c_g(b) \cdot b \cdot \delta(b),\]
where $b$ is the size of all dense chunks in $f$ and $g$. By definition,
$c_f(n)=1$, and $c_g(n) \leq m/n$, so we know that
$c_f(n)\, c_g(n)\, n\, \delta(n) \leq m\,\delta(n)$.
Because the chunk sizes $d_i$ were originally chosen by
Algorithm~\ref{alg:findk}, we must therefore have
$m\sum_{i=1}^t \delta(d_i) \leq m\delta(n)$. The restriction that the
$\delta$ function grows more slowly than linear then implies that
$(m/r)\sum \delta(d_i/s) \in O((m/r) \delta(n/s))$, and so the standard
equal-spaced algorithm is never more efficient in this case.

When $m\leq n$,
the number 
of ring operations to compute the product using the combined method, 
again by Theorem~\ref{thm:eqmul},
is
\begin{equation}\label{eqn:combcost}
	O\left( \delta(m/s) \sum_{d_i \geq m} (d_i/r) + 
	(m/r) \sum_{d_i < m} \delta(d_i/s) \right),\end{equation}
compared with $O((n/r)\delta(m/s))$ for the standard equal-spaced
algorithm. Because we always have $\sum_{i=1}^t d_i \leq n$, the first
term of \eqref{eqn:combcost} 
is $O((n/r)\delta(m/s))$. Using again the inequality
$m\sum_{i=1}^t \delta(d_i) \leq m\delta(n)$, along with the fact that
$m\delta(n) \in O(n\delta(m))$ because $m\leq n$, we see that the second
term of \eqref{eqn:combcost} is also $O((n/r)\delta(m/s))$. Therefore
the cost of the combined method is never more than the cost of 
equal-spaced multiplication alone.
\end{pf}

\section{Conclusions and Future Work}
\label{sec:conclusion}

Two methods for adaptive polynomial multiplication have been given where we
can compute optimal representations (under some set of restrictions)
in linear time in the size of the input. 
Combining these two ideas into one algorithm inherently captures
both measures of difficulty, and will in fact have significantly better
performance than either the chunky or equal-spaced algorithm in many
cases.

However, converting a sparse polynomial to the equal-spaced
representation in linear time is still out of reach, and this problem
is the source of the restriction of Theorem~\ref{thm:eqspchunks}. 
Some justification for the
impossibility of such a conversion algorithm was given, due to the fact
that the exponents could be long integers. However, we still do not have
an algorithm for sparse polynomial to equal-spaced conversion under the
(probably reasonable) restriction that all exponents be word-sized
integers. A linear-time algorithm for this problem would be useful and
would make our adaptive approach more complete, though slightly more
restricted in scope.

Some early results from a trial implementation indicate that the
algorithms we present are quite good at computing efficient adaptive
representations, even in the presence of ``noise'' in the input
polynomials, and although the conversion does sometimes have a
measurable cost, it is almost always significantly less than the cost of
the actual multiplication. 
Some of these results were reported in
\citep{Roc08}, giving evidence that our theoretical results hold in practice,
but more work on an efficient
implementation is still needed.

Yet another area for further development is multivariate polynomials.
We have mentioned the usefulness of Kronecker substitution, but developing
an adaptive algorithm to choose the optimal variable ordering would give
significant improvements.

Finally, even though we have proven that our algorithms produce optimal
adaptive representations, it is always under some restriction of the way
that choice is made (for example, requiring to choose an ``optimal
chunk size'' $k$ first, and then compute optimal conversions given $k$).
These results would be significantly strengthened by proving lower
bounds over all available adaptive representations of a certain type,
but such results have thus far been elusive.

\begin{ack}
The original ideas for this work were hatched in a graduate seminar taught
by Alex L\'opez-Ortiz and J\'er\'emy Barbay at the University of Waterloo.
Many thanks are due to the author's Ph.D. supervisors, Mark Giesbrecht
and Arne Storjohann, for their intellectual and financial support.
Thanks also to Richard Fateman and Michael Monagan for useful and
stimulating discussions on this work.
\end{ack}